\documentstyle[12pt]{article}

\newcommand{\be}{\begin{equation}}
\newcommand{\ee}{\end{equation}}
\newcommand{\bea}{\begin{eqnarray}}
\newcommand{\eea}{\end{eqnarray}}

\newcommand{\im}{\mbox{i}}

\begin{document}

\begin{titlepage}
\begin{flushright}
Freiburg THEP-97/1\\
gr-qc/9701018
\end{flushright}

\vspace{2cm}

\begin{center}

{\large\bf  SEMICLASSICAL QUANTUM STATES FOR
            BLACK HOLES}\footnote{Dedicated to Tullio Regge
            on his sixty--fifth birthday. To appear in the 
            Proceedings of the Second Conference on Constrained
            Dynamics and Quantum Gravity, Santa Margherita, 
            Italy, September~1996, edited by V. de~Alfaro {\em et al.}
            (Nucl. Phys.~B, Suppl., 1997).}
\vskip 1cm
{\bf Claus Kiefer}
\vskip 0.4cm
 Fakult\"at f\"ur Physik, Universit\"at Freiburg,\\
  Hermann-Herder-Stra\ss e 3, D-79104 Freiburg, Germany.

\end{center}
\vskip 2cm
\begin{center}
{\bf Abstract}
\end{center}
\begin{quote}
I discuss the semiclassical approximation for the Wheeler--DeWitt
equation when applied to the CGHS model and spherically symmetric
gravity. Special attention is devoted to the issues of
Hawking radiation, decoherence of semiclassical states, and
black hole entropy.
 \end{quote}

\end{titlepage}

One of the most important unsolved problems
in theoretical physics is to understand
the full non--perturbative evolution of a black hole. Since a
consistent quantum theory of gravity does not seem available
yet, such an understanding remains elusive. A quantum theory of
gravity is needed in particular to provide a derivation of black hole
entropy from quantum statistical mechanics. 

As long as the full theory is unknown, research is focused
on approaches which are either candidates for such a theory,
or probably provide interesting models within which some of
the interesting questions can be addressed. One of these approaches
is canonical quantum gravity. Since the full Wheeler--DeWitt equation
cannot be tackled in general, a restriction to special models
and approximation schemes is necessary. In the following I shall
briefly discuss some results which can be obtained from 
semiclassical approximations. I shall assume that there are no
anomalies present. Otherwise the scheme has to be modified
\cite{CJZ}. Most results are in accordance with results found
by other methods, but it is illuminating to view them from
a different perspective. 
I shall restrict attention in
the following to basically two aspects: First, the treatment
of Hawking radiation in the functional Schr\"odinger equation
and its decohering influence on black hole superpositions and,
second, the notion of black hole entropy for semiclassical
quantum states. Most of the material was developed in collaboration
with Jean-Guy Demers \cite{DK} and Thorsten Brotz \cite{BK},
and I refer to these papers for more details.

The semiclassical approximation to canonical quantum gravity,
in its simplest version, is performed through an expansion with
respect to the gravitational constant $G$ \cite{Ki}.
Starting from the full quantum equations (Wheeler-DeWitt equation
and diffeomorphism constraints)
\be \hat{\cal H}_{\perp}\Psi=0, \;\;\;
    \hat{\cal H}_i\Psi=0 \ee
for a wave functional $\Psi$ depending on three--geometries
(as represented by three--metrics) and matter fields,
an ansatz of the form
\be \Psi=\exp \im\left(G^{-1}S_0+S_1+GS_2+\ldots\right) \ee
yields equations at consecutive orders of $G$:
At order $G^{-1}$ the gravitational Hamilton--Jacobi
equation is found and the notion of a background spacetime
recovered. The next order ($G^0$) yields a functional Schr\"odinger
equation for non--gravitational fields in such a background
spacetime. Higher order terms lead to genuine quantum gravitational
corrections \cite{Ki}, but they will not be discussed here.

Two specific models will be addressed. In the first, the gravitational
part will consist of the metric and a dilaton field $\phi$ in one
space dimension.
It is known as the CGHS model \cite{CGH}. The two--dimensional
line element reads
\be ds^2= e^{2\rho}\left(-N^2dt^2+ [dx+N_xdt]^2\right), \ee
where $N$ and $N_x$ are the rescaled lapse and shift function,
respectively. The gravitational configuration space thus consists
of the fields $\rho(x)$ and $\phi(x)$.

 The second model is spherically symmetric gravity
in $3+1$ dimensions. The line element is here chosen to read
\bea ds^2 &=& -N^2dt^2 +\Lambda^2(r,t)(dr+N_rdt)^2
      \nonumber\\ & & \ +R^2(r,t)
    d\Omega_2^2, \eea
where $N$ and $N_r$ are here the original lapse and shift function,
respectively, and $d\Omega_2^2$ denotes the line element on the
two--sphere. The gravitational configuration space here thus
consists of the fields $\Lambda(r)$ and $R(r)$.
 These two models can actually be considered as special
cases of a more general class of models, but for the following
discussion they will be treated separately.

The action of the CGHS model reads \cite{CGH}
\be S=\int dxdt\sqrt{-g}\left[\frac{1}{G}(R\phi+4\lambda^2)
    -\frac{1}{2}(\nabla f)^2\right], \ee
where a massless scalar field $f$ has been added. (Strictly
speaking, the variable $\rho$ and $\phi$ which occur in this action
have been redefined to cancel the kinetic term for the dilaton,
see \cite{DK}.)
The quantum equations (1) can be derived from this action in the
standard canonical way. As mentioned above,
 order $G^{-1}$ of the
semiclassical approximation with respect to the 
gravitational constant (which is here dimensionless)
 yields the Hamilton-Jacobi equation
for $\rho$ and $\phi$. A one--parametric class of
 solutions reads \cite{LM}
\be S_0= \int dx\left[Q+\phi'\ln\left(\frac{2\phi'-Q}
         {2\phi'+Q}\right)\right], \ee
where
\be Q\equiv 2\sqrt{\phi'^2+[C-4\lambda^2\phi]e^{2\rho}}. \ee
{}From $S_0$ one can recover classical black hole solutions
(``eternal holes") with an ADM mass $M\equiv C/4\lambda$, 
in accordance with the spacetime treatment presented in \cite{CGH}.
There also exists a solution describing a collapse towards a
black hole, if part of the $f$-field is in a semiclassical
state corresponding to an incoming shock wave.

In the next order ($G^0$), the full wave functional assumes
the approximate form
\be \Psi\approx e^{\im G^{-1}S_0[\rho,\phi]}
     \chi[\rho,\phi,f] , \ee
where $\chi$ obeys the functional Schr\"odinger equation
\be \im\frac{\partial\chi}{\partial t} =
    \frac{1}{2}\int dx\left(-\frac{\delta^2}{\delta f^2}
     +f'^2\right)\chi. \ee
(The semiclassical time parameter $t$ is found from
$\partial/\partial t=\nabla S_0\cdot\nabla$.)
As a consequence of the conformal nature of the coupling
in two dimensions, this is just a free Schr\"odinger equation.
How, then, can the Hawking radiation be recovered?
The answer to this question lies in the proper handling
of {\em boundary conditions}.

 Consider the semiclassical solution
where the $f$-shock wave produces a black hole.   
The idea is to start with the vacuum state at early times,
when no black hole is present. A ``vacuum state" is here always defined
with respect to coordinates which exhibit the asymptotic flatness
of the metric (``inertial coordinates"). Such vacuum states are
represented by Gaussian wave functionals. With these inertial
coordinates, the functional Schr\"odinger equation (9) is then
solved in the black hole region. The slices are chosen in such
a way that they pass through the bifurcation point.
They thus do not penetrate the event horizon.
The resulting Gaussian functional does not, however, correspond
to a vacuum state for an observer in the presence of the hole.

The normalisation of the wave functional demands the boundary
condition that the field $f$ go to zero as the bifurcation point is
approached \cite{DK}. This mimicks the presence of a {\em mirror}
in an analogous discussion for accelerated observers in
Minkowski space \cite{Fr}. The explicit form of the solution
to (9) reads (where $f(k)$ is the Fourier transform of
$f(x)$)
\bea & & \chi[f(k),t)= N\times \nonumber\\
       & & \exp\left(-\int_{-\infty}^{\infty}
     dk\ k\coth\left(\frac{\pi k}{2\lambda}\right)
     \vert f(k)\vert^2\right). \eea
In the language of quantum optics, this is just a squeezed vacuum state.
If the expectation value of the number operator $a^{\dagger}a$
for a mode with wave number $k$ is performed (where $a$ and 
$a^{\dagger}$ are defined with respect to the vacuum state in the
presence of the hole), one finds a Planckian distribution
with respect to the Hawking temperature $T_H=\lambda/2\pi$.
Still, the state (10) is a pure state and can be distinguished
from a canonical ensemble by performing expectation values
of other operators.

With the above choice of slicing, the quantum correlations
between exterior and interior region of the hole are not captured.
These correlations can be taken into account, if -- in spite of the
problems with the normalisation -- the boundary condition 
$f\to 0$ at the bifurcation point is relaxed. Tracing out the
degrees of freedom in the second, inaccessible, exterior region
leads then to a density matrix with respect to the above
Hawking temperature in the black hole region, in analogy to
\cite{Fr,Is}. 

The above discussion has been made for a state of the form (8).
Since the equations (1) are linear, arbitrary superpositions
of such states are expected to occur. Such superpositions do not
lead to a semiclassical background. Nor do they yield a
Schr\"odinger equation at order $G^0$. The physical mechanism
of decoherence \cite{deco} can, however, explain why the various
semiclassical components can be considered as dynamically independent.
Interference terms are delocalised by correlations with the 
Hawking radiation. This was shown explicitly in \cite{DK}
for a superposition of a black hole with a white hole,
and for a superposition of a black hole with no hole.
{}From a physical point of view, this is just an effect of
symmetry breaking (see Chap.~9 of \cite{deco}).
If the holes are put in a box, decoherence is efficient if the size
of the box is much bigger than the dominating wavelength of the
radiation.

I shall now turn to a discussion of {\em entropy} in the context
of semiclassical quantum states. Here, the line element (4)
of spherically symmetric gravity in $3+1$ dimensions is chosen,
and a spherically symmetric electromagnetic potential is
added \cite{BK}. (The following analysis can also be made
for the CGHS model, while the discussion above cannot be extended
in a straightforward way to spherically symmetric gravity.) 
 For the potential the ansatz $A=\varphi dt+\Gamma dr$
is made, with $\Gamma$ playing the role of the canonical variable.
Making use of the electromagnetic Gauss constraint, the full wave
functional can be written in the form
\be \Psi= \exp\left(iq\int_{-\infty}^{\infty}\Gamma(r)dr\right)
        \psi[\Lambda(r),R(r)], \ee
where $\psi$ satisfies an effective Wheeler-DeWitt equation.
The semiclassical approximation is performed for the latter.
The solution to the Hamilton-Jacobi equation looks similarly to (6)
and depends only on two parameters: the mass $m$ and the charge $q$
of the hole. (This demonstrates that this model is classically
equivalent to a finite--dimensional model.)

How can the Bekenstein-Hawking entropy be recovered from such states?
A key observation in this respect is the fact that boundary terms
have to be considered in gravitational actions.
A useful analogy is the case of {\em asymptotically flat}
three--geometries which leads to the presence of {\em additional}
degrees of freedom, in particular the ADM mass \cite{RT}.
If one chooses in the present case slices which originate at the 
bifurcation point, boundary conditions there lead to
{\em new degrees of freedom} in the canonical formalism
\cite{BK,CT}: the area $A$ of the horizon and its conjugate variable
(the ``deficit angle").

A pure state has of course vanishing entropy. The Bekenstein-Hawking
entropy can thus only be recovered if the standard transition
to the euclidean formalism is performed \cite{Ha}.
There is then a close connection between euclidean quantum states
and partition sums. In the present case the euclidean states read
\be \psi_E\approx\exp\left(-S_0^E-\beta m+\frac{A}{4G}\right), \ee
where $S_0^E$ is the euclidean version of the Hamilton-Jacobi
functional, and $\beta$ is the inverse temperature at spatial infinity.
One recognises the explicit occurrence of the Bekenstein-Hawking
entropy $A/4G$ in (12).

On the semiclassical level -- and after decoherence between
states of the form (8) is taken into account --
it still makes sense to talk of {\em slices} through a background
spacetime. The entropy discussed above then depends on the
choice of these slices. If they start at the bifurcation point,
only the exterior part can be recovered from initial data on
these slices. The unavailable information about the interior
region is then expressed by the occurrence of the entropy in (12).
If a slice is chosen that originates at the singularity in the
Schwarzschild case, more information is available, and one expects
that the entropy be lower. This is in fact what turns out to be the 
case: for such slices the entropy is only one fourth of the
Bekenstein-Hawking entropy \cite{BK}.
A similar interpretation was given from a path integral point
of view in \cite{Ma}.

For the extremal Reissner-Nordstr\"om black hole, it turns out
that no additional degrees of freedom arise at the bifurcation point.
This fits also nicely
into the above picture: for nonextreme holes one can only recover
part of the spacetime from slices originating at the bifurcation
point, while for the extreme case the whole spacetime
(up to the Cauchy horizon) can be constructed. This is
straightforward to see by an inspection of the corresponding
Penrose diagrammes. 

While this result of vanishing entropy for extreme holes is
in accordance with semiclassical results from the path integral
\cite{Sex}, it is in disagreement with recent results from
string theory \cite{Ho}: While there exists an extremal hole
with vanishing entropy, it turns out to be unstable, while
the stable one has the Bekenstein-Hawking value for the entropy.

In the path integral language, the above quantum state (12)
corresponds to the ``tree level" of approximation.
It would be interesting to discuss the entropy at the level
of the next (``one loop") order, corresponding to (8).
The contributions to entropy at this order should yield both
the entropy of the field outside the horizon as well as
corrections to the Bekenstein-Hawking entropy \cite{Fro}.
This, however, will be discussed elsewhere.  

\vskip 2mm

I thank T.~Brotz for a critical reading of this paper.

\end{document}